\documentclass[reprint, superscriptaddress, amsmath,amssymb, aps,]{revtex4-2}

\usepackage{graphicx}
\usepackage{dcolumn}

\usepackage{lipsum} 
\usepackage{xcolor}

\newcommand{\ins}{\text{in}}
\newcommand{\out}{\text{out}}

\newcommand{\dil}{\text{dil}}
\newcommand{\den}{\text{den}}
\newcommand{\core}{\text{core}}

\newcommand{\eqden}{c^\text{den}}
\newcommand{\eqdil}{c^\text{dil}}

\newcommand{\eqdelta}{\Delta c}

\newcommand{\oi}{\text{out/in}}
\newcommand{\dilden}{\text{dil/den}}

\newcommand{\Rio}{\text{i/o}}
\newcommand{\Ri}{\text{i}}
\newcommand{\Ro}{\text{o}}

\newcommand{\eq}[1]{Eq.~(\ref{eq:#1})}
\newcommand{\fig}[1]{Fig.~\ref{fig:#1}}

\newcommand{\qd}{\quad .}
\newcommand{\qc}{\quad ,}

\newcommand{\nn}{\nonumber}

\usepackage{xr}
\begin{document}

\title{
Formation of liquid shells in active droplet systems
}

\author{Jonathan Bauermann}
\affiliation{
Max Planck Institute for the Physics of Complex Systems,
 Nöthnitzer Stra\ss e~38, 01187 Dresden, Germany
}
\affiliation{
Center for Systems Biology Dresden,  Pfotenhauerstra\ss e~108, 01307 Dresden, Germany
}
\author{Giacomo Bartolucci}
\affiliation{
Max Planck Institute for the Physics of Complex Systems,
 Nöthnitzer Stra\ss e~38, 01187 Dresden, Germany
}
\affiliation{
Center for Systems Biology Dresden,  Pfotenhauerstra\ss e~108, 01307 Dresden, Germany
}
\affiliation{
 Faculty of Mathematics, Natural Sciences, and Materials Engineering: Institute of Physics, University of Augsburg, Universit\"atsstra\ss e~1, 86159 Augsburg, Germany
}
\author{Job Boekhoven}
\affiliation{
School of Natural Sciences, Department of Chemistry, Technical University of Munich, Lichtenbergstra\ss e~4, 85748 Garching, Germany
}

\author{Christoph A. Weber}
\email{christoph.weber@physik.uni-augsburg.de}
\affiliation{
 Faculty of Mathematics, Natural Sciences, and Materials Engineering: Institute of Physics, University of Augsburg, Universit\"atsstra\ss e~1, 86159 Augsburg, Germany
}
\author{Frank Jülicher}
\email{julicher@pks.mpg.de}
\affiliation{
Max Planck Institute for the Physics of Complex Systems,
 Nöthnitzer Stra\ss e~38, 01187 Dresden, Germany
}
\affiliation{
Center for Systems Biology Dresden,  Pfotenhauerstra\ss e~108, 01307 Dresden, Germany
}
\affiliation{
Cluster of Excellence Physics of Life, TU Dresden, 01062 Dresden, Germany
}

\date{\today}

\begin{abstract}
We study a chemically active binary mixture undergoing phase separation and show that under non-equilibrium conditions, stable liquid spherical shells can form via a spinodal instability in the droplet center. A single liquid shell tends to grow until it undergoes a shape instability beyond a critical size.
In an active emulsion, many stable and stationary liquid shells can coexist.
We discuss conditions under which liquid shells are stable and dominant as compared to regimes where droplets undergo shape instabilities and divide. 
\end{abstract}

\maketitle
Emulsions are heterogeneous liquids where phase separation leads to the formation of droplets that coexist with the surrounding fluid.
They play an important role in many fields, from physics and chemistry to biology and engineering, for examples see~\cite{lohse:2020,aulton:2013,mcclements:2015}.
In living cells, proteins and nucleic acids often condense together into assemblies with properties of liquid-like droplets~\cite{brangwynne:2009, brangwynne:2011, weber:2012, hyman:2014, alberti:2017,boeynaems:2018, choi:2020, fritsch:2021}. The cell interior, therefore, resembles an emulsion. Because biochemical processes operate away from thermodynamic equilibrium, this emulsion is inherently active, driven by energy input at the molecular scale.
In this active environment, liquid condensates in cells provide biochemical compartments that play a role in the spatial organization of biological processes~\cite{molliex:2015,banani:2017,shin:2017a, weber:2019a, nakashima:2019,donau:2023}.

The role of liquid condensates in cells suggests that phase separation can more generally play a role in the organization of chemical reactions in many systems.
Phase-separated systems are not dilute, and therefore mass-action laws for chemical kinetics have to operate in a non-dilute regime which leads to an interplay between phase separation physics and reaction chemistry~\cite{bauermann:2022c,zwicker:2022}.
Here, we are interested in how phase separation can spatially organize chemical processes and how chemical activity influences the morphology of phase-separated droplets.
Previous theoretical work has shown interesting consequences of chemical activity on phase-separated droplets. For example, Ostwald ripening can be suppressed in an active emulsion because diffusion coefficients and reaction rates introduce characteristic length scales~\cite{glotzer:1994a, glotzer:1995, zwicker:2015, wurtz:2018}. 
Ostwald ripening can also be accelerated if chemical reactions act on the gradients between droplets~\cite{tena-solsona:2021}.
Furthermore, it was shown that for sufficiently strong non-equilibrium driving spherical droplets can undergo a shape instability that, in three dimensions, can give rise to spontaneous droplet division and the emergence of growth and division cycles~\cite{zwicker:2017a, li:2020a, bauermann:2022d, demarchi:2023}.

In recent work, synthetic analogs of such chemically active emulsions have been developed~\cite{donau:2020a, yewdall:2021, spath:2021, donau:2022}.
In such a synthetic chemically active emulsion, it was recently shown that stable liquid shells can form as non-equilibrium structures in systems where two phases coexist~\cite{bergmann:2023}.
In this case, a liquid shell forms one phase that coexists both on the inside and the outside with a second phase. 
In numerical studies, ring-like patterns have been reported as stationary states in a ternary mixture with active molecular transitions~\cite{bartolucci:2021a}.
In passive systems, spherical shells can emerge from three-phase coexistence~\cite{gibbs:1876,mao:2019a,mao:2020,spruijt:2023}. This situation arises when a droplet of phase I is located inside a droplet of phase II that itself is immersed in phase~III.

In this paper, we study a simple model of a binary fluid composed of components $A$ and $B$ that undergoes phase separation in an $A$-rich and a $B$-rich phase and which takes into account chemical transition between $A$ and $B$ molecules. 
Because we are interested in active systems that are driven out of equilibrium by an energy input in the chemical reactions, these transitions do not obey a detailed balance condition.
We use a two-phase system as a minimal model to investigate conditions under which liquid shells emerge spontaneously.
We discuss the growth dynamics of individual shells and show that they can be stable and stationary in an active emulsion, where they can coexist with other shells and droplets, both in two and three dimensions.
We also discuss parameter regimes where spherical shells emerge,
and compare them to regimes where droplets exist. Finally, we discuss shape instabilities of growing shells and put them in relation to shape instabilities of droplets.

\begin{figure}
\centering
\includegraphics[width=0.5\textwidth]{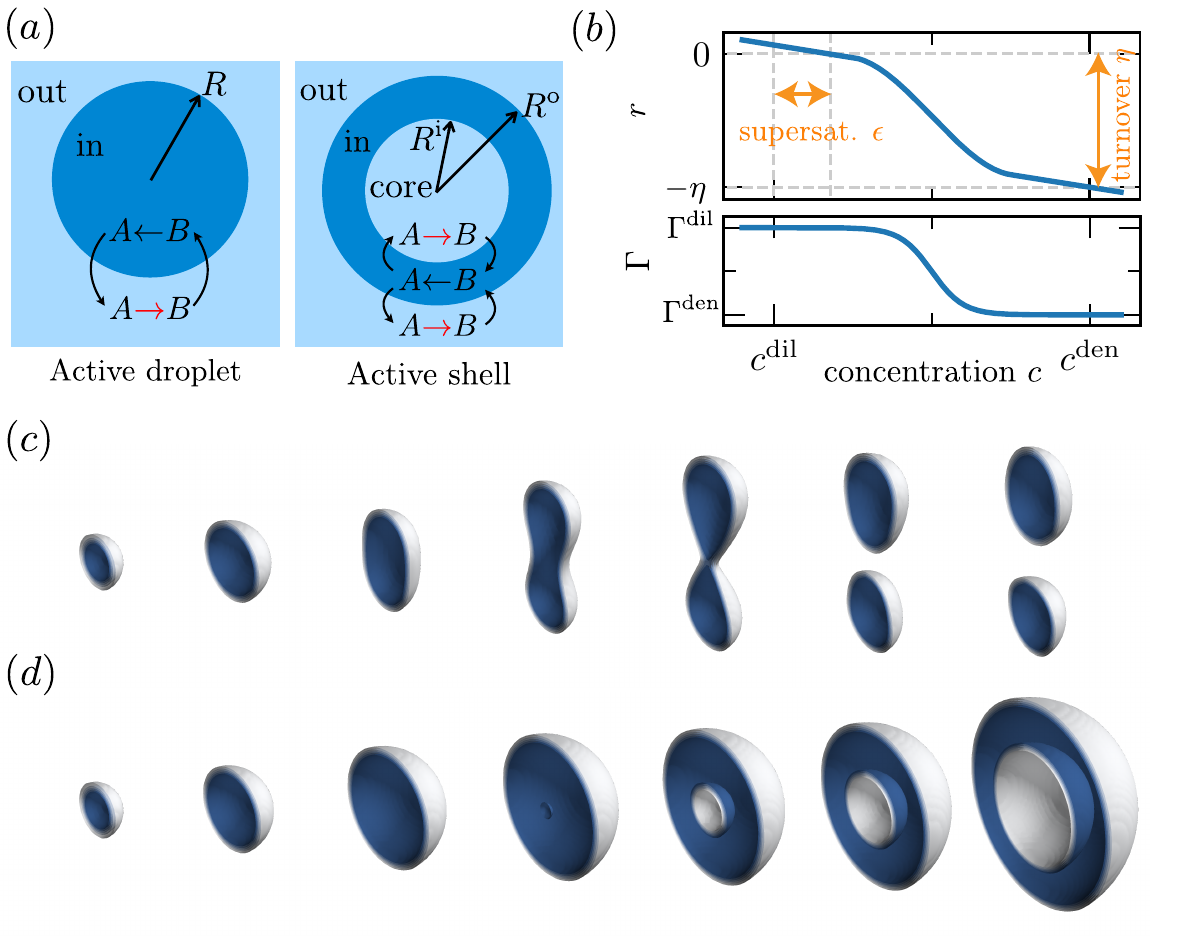}
\caption{
(a) Minimal model of chemically active droplets (dilute/dense phase in the outside/inside domain) and shells (dilute/dense phase in the outside \& core/inside domain).
(b) Concentration dependency of the reaction flux $r$ and the mobility $\Gamma$ (App.~\ref{app:mobility}). 
(c-d) Cuts through temporal consecutive three-dimensional contour surfaces of a droplet (c) and a shell (d) with
$b=\kappa=1$, $c^\dil=0$, $c^\den=1$, $\Gamma^\dilden=1$,  $\eta=0.003$, $\epsilon=0.19$, $k^\dilden=0.0025$ (c), $k^\dilden=0.01$ (d). For details, see App.~\ref{app:numerics}.
\label{fig:model}
}
\end{figure}

\emph{Minimal model of chemically active droplets} - 
We consider an incompressible binary fluid described by local thermodynamics governed by the chemical potential $\mu(\mathbf{x})=\delta F/ \delta c(\mathbf{x})$, where 
\begin{equation}
\label{eq:free_ener}
F = \int dV \left( f(c) + \frac{\kappa}{2} (\nabla c)^2\right) \qc
\end{equation}
is the free energy. Here, $c(\mathbf{x})$ denotes the concentration field of the $B$ component, $\mathbf{x}$ is a position in space, $dV$ the volume element in two or three dimensions, and $\kappa$ a parameter describing the free energy contribution of concentration gradients.
We choose the free energy density~\cite{ginzburg:2009} $f(c) = {b (c-\eqdil)^2(c-\eqden)^2} /(2 \eqdelta^2)$, where $\eqdelta=\eqden-\eqdil$, and $\eqdil$ and $\eqden$ are the equilibrium concentrations of the coexisting dilute and dense phases, respectively.
The parameter $b$ describes the strength of segregation. The surface tension of the interface is given by  $\gamma =\eqdelta^2 \sqrt{ \kappa b} /6$~\cite{degennes:1985a}.

The time evolution of the concentration field is a Cahn-Hilliard equation \cite{cahn:1961} complimented by reactions, which reads
\begin{gather}
\label{eq:dyn_eq}
    \partial_t c = - \nabla \cdot \mathbf{j} + r(c) \qc  \\
    \mathbf{j} = -\Gamma(c) \nabla \mu(c) \qc \label{eq:dyn_flux}
\end{gather}
where $\mathbf{j}$ is a diffusive flux, driven by a gradient of the chemical potential $\mu$ and proportional to the mobility $\Gamma(c)$, which in the dilute and the dense phase takes the values $\Gamma^\dil$  and $\Gamma^\den$, respectively with a smooth concentration dependency across the interface, see \fig{model}(b). 
The reaction flux is described by first-order reaction kinetics with different coefficients in the two phases.
The linear pieces are smoothly connected by a third order polynomial $p(c)$ (see \fig{model}(b)):
\begin{align}
\label{eq:reac_flux}
    r(c) =
\begin{cases}
-k^\dil c + s^\dil, &c \leq c^\dil_\text{B}\\
p(c), & c^\dil_\text{B}<\phi<c^\den_\text{B}\\
-k^\den c + s^\den, & c \geq c^\den_ \text{B}
\end{cases}\qd
\end{align}
This reaction flux defines two important parameters, the supersaturation in the dilute phase far from the droplet $\epsilon = s^\dil/k^\dil - \phi_0^\dil$ and the turnover rate $\eta = k^\den \eqden - s^\den $, which describes the reaction flux in the dense phase.
Note that for phase coexistence, linear reaction fluxes are not consistent with thermodynamic equilibrium, and our choice corresponds to an active reaction coupled to an external fuel reservoir \cite{bauermann:2022c}. 

For turnover rates $\eta>0$ and supersaturations $\epsilon$ larger than a critical value, spherical droplets of finite radius $R$ exist as steady-state solutions of Eqs.~\eqref{eq:dyn_eq} and \eqref{eq:dyn_flux} \cite{zwicker:2017a}. If $\epsilon$ is increased further, the spherical state becomes unstable, which in three dimensions typically leads to droplet division, see \fig{model}(c).
If instead of increasing $\epsilon$ the reaction rate $k^{\rm den}$ is increased, a spherical shell can form via a spinodal instability inside the droplet, see \fig{model}(d).
This instability occurs because, inside the droplet, the dense phase concentration drops below the spinodal concentration. This concentration drop results from a steeper concentration profile inside the droplet for faster material turnover.
After this instability, a spherical shell of the dense phase emerges, which can be characterized by the radii $R^\Rio$, which both increase with time.

\emph{Quantitative analysis in a sharp interface limit} - 
We perform a detailed stability analysis of individual droplets and shells in a sharp interface limit in spherical coordinates with radial coordinate $r$
and angular coordinates $\varphi$ and $\theta$. 
In the vicinity of the equilibrium concentrations $c_0^\dilden$ that coexist at a flat interface, we linearize the dynamic equations Eqs.~\eqref{eq:dyn_eq} and \eqref{eq:dyn_flux} in the dense and dilute phase. For a droplet, we write
\begin{equation}
    \partial_t c^i = D^i \nabla^2 c^i - k^i c^i + s^i \label{eq:reac-dif}
\end{equation}
where $i={\rm out, in}$ describe the dilute phase in the outside domain and the dense phase in the inside domain of the droplet, respectively. We find the diffusion coefficients 
$D^\oi=\Gamma^\dilden b$, the reaction rates $k^\oi=k^\dilden$, and the source rates $s^\oi=s^\dilden$. The droplet shape is described by the radial position $r=R(\theta,\varphi)$ of the interface.
A spherical shell is also described using Eq.~\eqref{eq:reac-dif} but with three domains $i={\rm out,in,core}$, a dilute phase outside domain, a dense phase inside the shell and a dilute phase in the core domain with $D^\core=\Gamma^\dil b$, $k^\core=k^\dil$, and $s^\core=s^\dil$. The three domains are separated by two interfaces $r=R^\Rio(\theta,\varphi)$, see \fig{model}(a).

The boundary conditions at an interface follow from local thermodynamic equilibrium~\cite{langer:1975}. At an interface with local mean curvature $H$, the coexisting concentrations are given by the Gibbs-Thompson relation
\begin{equation} 
\label{eq:bc_inter}
c^\oi = c^\dilden + \beta \gamma H \qc
\end{equation}
where $\beta=1/(b \Delta c)$ describes the curvature dependence of equilibrium concentrations~\cite{bray:1994}.
Here, the curvature is defined relative to a surface normal that points from the dense to the dilute phase.
In addition to these concentration boundary conditions, material is conserved when diffusing across the interface. This condition specifies the normal interface velocity
\begin{equation}
\label{eq:drop_vel}
    \frac{dR}{dt} = \frac{\mathbf{j}^i(R) - \mathbf{j}^j(R)}{c^i(R) - c^j(R)} \cdot \mathbf{e}_r 
\end{equation}
of the interface separating the domains $i$ and $j$~\cite{bray:1994}.
Here, $\mathbf{j}^i = -D^i \nabla c^i$ is the diffusive flux at the interface in domain $i$.

\begin{figure}
\centering
\includegraphics[width=0.49\textwidth]{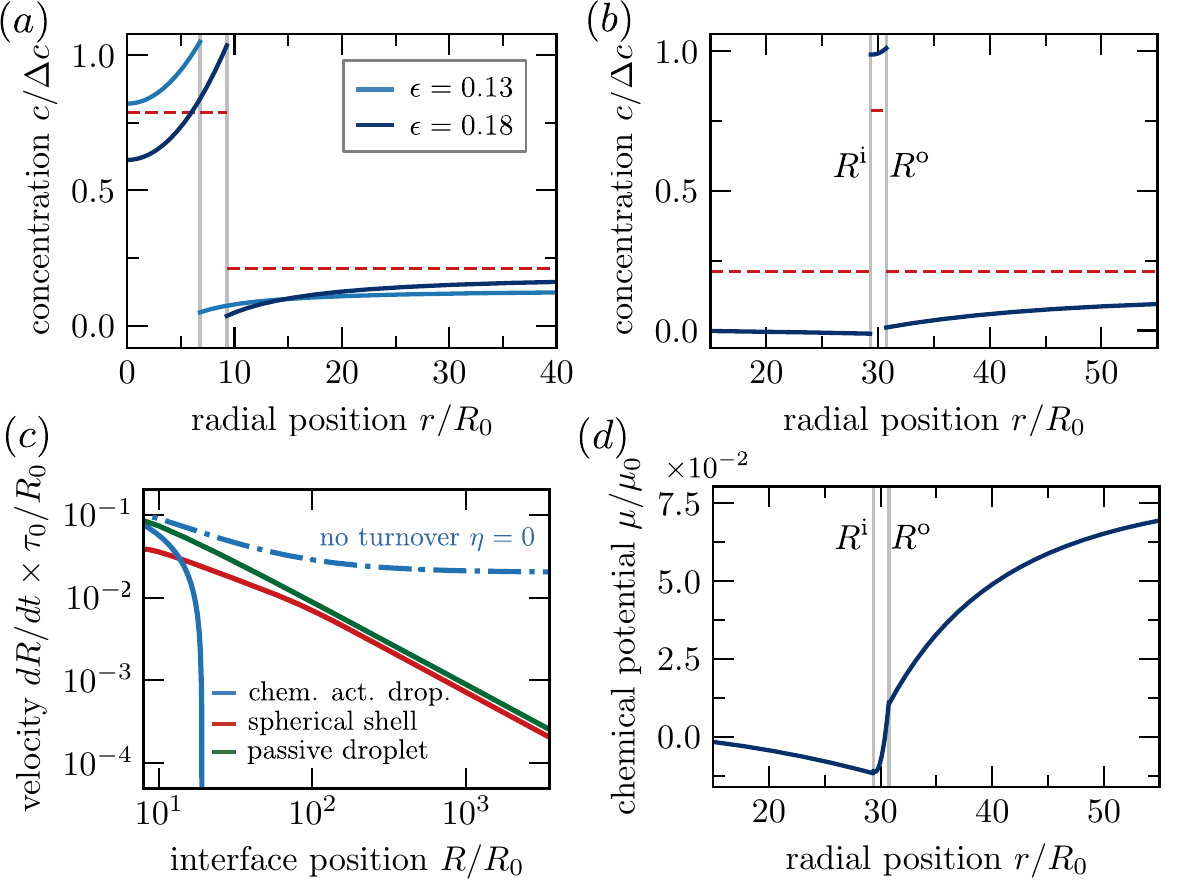}
\caption{
(a-b) Radial concentration profiles (blue lines) of stationary droplets (a) for $\epsilon=0.13,0.18$ and a stationary shell (b) for $\epsilon=0.18$ in a finite system with a no-flux boundary condition at $R_s/R_0=80$, the spinodal concentrations (red dashed lines), and interface positions (grey lines).
(c) Interface velocities for chemically active droplets (blue, solid line: $\epsilon=0.18$, $\eta=0.04$, dashed-dotted: $\epsilon=0.18$, $\eta=0$), and spherical shells (red, $\epsilon=0.18$, $\eta=0.04$), and a passive droplet in a supersaturated environment (green, $c^\infty=0.18$, $\eta=0$, $\epsilon=0$, $k^\dilden=0$) in an infinite system.
(d) Chemical potential profile for the shell shown in (b).
If not mentioned otherwise parameter are: $c^\dil=0$, $c^\den=1$, $D^\dil=1$, $D^\den=5$,  $k^\dilden=0.0025$;
Units are: $R_0=6 \beta \gamma /\Delta c$, $\tau_0 = R_0^2/D^\dil$, $\mu_0 = 1/(\beta \Delta c^3)$.
\label{fig:dropstab}}
\end{figure}

Spherical droplets of finite size $R$ are steady-state solutions of Eqs.~\eqref{eq:reac-dif} and \eqref{eq:drop_vel}~\cite{zwicker:2017a}, two examples of radial concentration profiles are shown in \fig{dropstab}(a). For $\epsilon=0.13$, the concentrations inside and outside remain in the stable regime above and below the spinodal concentrations, respectively  (red dashed line). 
For $\epsilon=0.18$, the concentration inside drops below the spinodal, indicating an instability of the droplet. As a consequence, a dilute phase is nucleated inside the droplet, forming the core domain, while the dense phase forms a spherical shell, see \fig{model}(a).
In a finite system, shells with stationary radii $R^\Rio$ can exist. A concentration profile of such a shell is shown in \fig{dropstab}(b). This solution is locally stable, i.e., outside the spinodal regime (red dashed line).
At the interfaces, the concentration jumps between dilute and dense phase.
Equation~(\ref{eq:bc_inter}) is obeyed at both radii $R^\Ri$ and $R^\Ro$ because $H$ has opposite signs at the two interfaces. Note that the chemical potentials vary continuously across the interfaces, see \fig{dropstab}(d).

\begin{figure}
\centering
\includegraphics[width=0.49\textwidth]{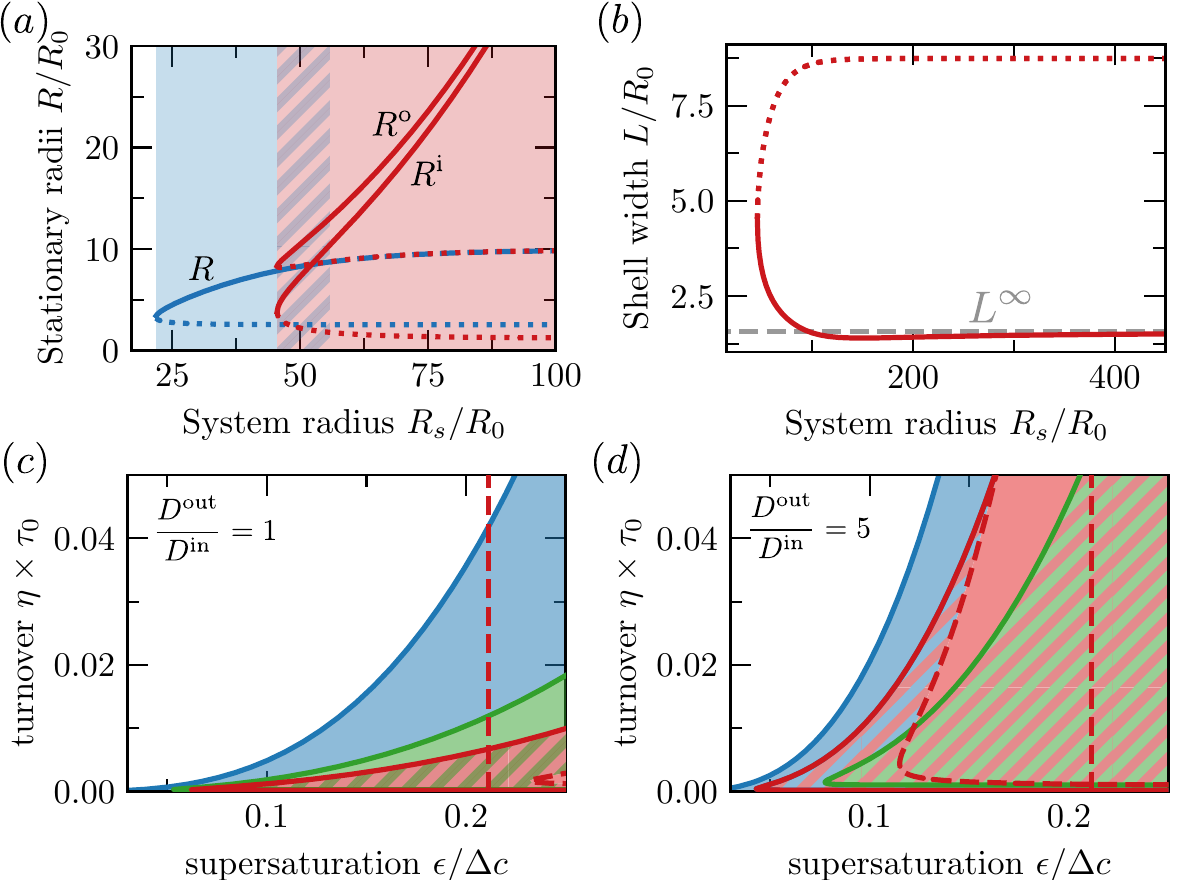}
\caption{
(a) Stationary interface positions for a droplets/shells in blue/red as a function of system size $R_s$. Stable (solid) $R$/$R^\Rio$ and critical (dotted) $R_c$/$R_c^\Rio$ droplet/shell interfaces exist. Droplet interfaces can undergo a spinodal instability (dotted).
(b) Interface width $L$ for shells in (a) as a function of system size $R_s$. 
(c-d) Stability diagrams for two different ratio of diffusivites of droplets and shells in a finite system ($R_s/R_0=80$). Blue region: droplets are stable, green region: droplets are shape unstable, red domain: shells are stable. Red dashed lines indicate the spinodal in the dilute (vertical line)  and droplet center (curved line). If not mentioned otherwise, parameters and units are the
same as in \fig{dropstab}.
\label{fig:vacstate}}
\end{figure}

To discuss time-dependent solutions for droplets and shells, we consider the case of slow interface velocities, where the concentration field is quasi-stationary. To this end, we determine stationary solutions of Eq.~\eqref{eq:reac-dif} with given interface positions and calculate the corresponding interface velocities $dR/dt$.
For an active droplet with $\eta>0$, the interface velocity vanishes at a stationary radius $R$, see \fig{dropstab}(c) (solid blue line).
If reactions inside the droplet are suppressed ($\eta=0$), the interface velocity approaches a constant value for large $R$ (dashed-dotted blue line).
A spherical shell exhibits a growth velocity that decreases for increasing radius $dR/dt \propto 1/R$ and does not reach a steady state in an infinite system.
This asymptotic behavior is the same as for a passive droplet in a supersaturated environment (solid green line).

We next discuss steady states with spherical symmetry in a finite system with radius $R_s$. Fig.~\ref{fig:vacstate}(a) shows the stationary interface radii of droplets and shells as a function of system size.
Beyond a minimal system radius $R_s$, stationary droplets with radii $R$ and $R_c$ exist. The radius $R$ can be stable (solid blue line), and the radius $R_c$ is the unstable critical radius (dotted blue line).
For larger $R_s$, stationary shells exist with inner and outer radii $R^\Rio$ and their critical counterparts $R_c^\Rio$ (solid and dotted red lines).
Stable droplets (shells) exist in the blue (red) shaded region. 
In the region shaded both in blue and red, stable solutions for droplets and shells coexist. 
The stable droplets undergo a spinodal instability at the end of the blue-shaded region. Beyond this point, only shells are stable (solid red line), while droplets are unstable (dashed blue line).
Note that stationary droplets reach a finite radius for large systems, while stationary shell radii grow proportional to system size.
Furthermore, the width $L=R^\Ro-R^\Ri$ of shells reaches in the large system limit the finite value
\begin{equation}
\label{eq:Linfty}
  L^\infty = 2 \sqrt{\frac{D^\den}{k^\den}}  \text{arctanh}
  \left(\sqrt{\frac{D^\dil k^\dil}{D^\den k^\den}} \frac{k^\den \epsilon} {\eta} \right) \qc
\end{equation}
which is indicated in \fig{vacstate}(b). This value equals the width of a stationary droplet size in a one-dimensional system, see App.~\ref{app:Linfty}.

\begin{figure}
\centering
\includegraphics[width=0.5\textwidth]{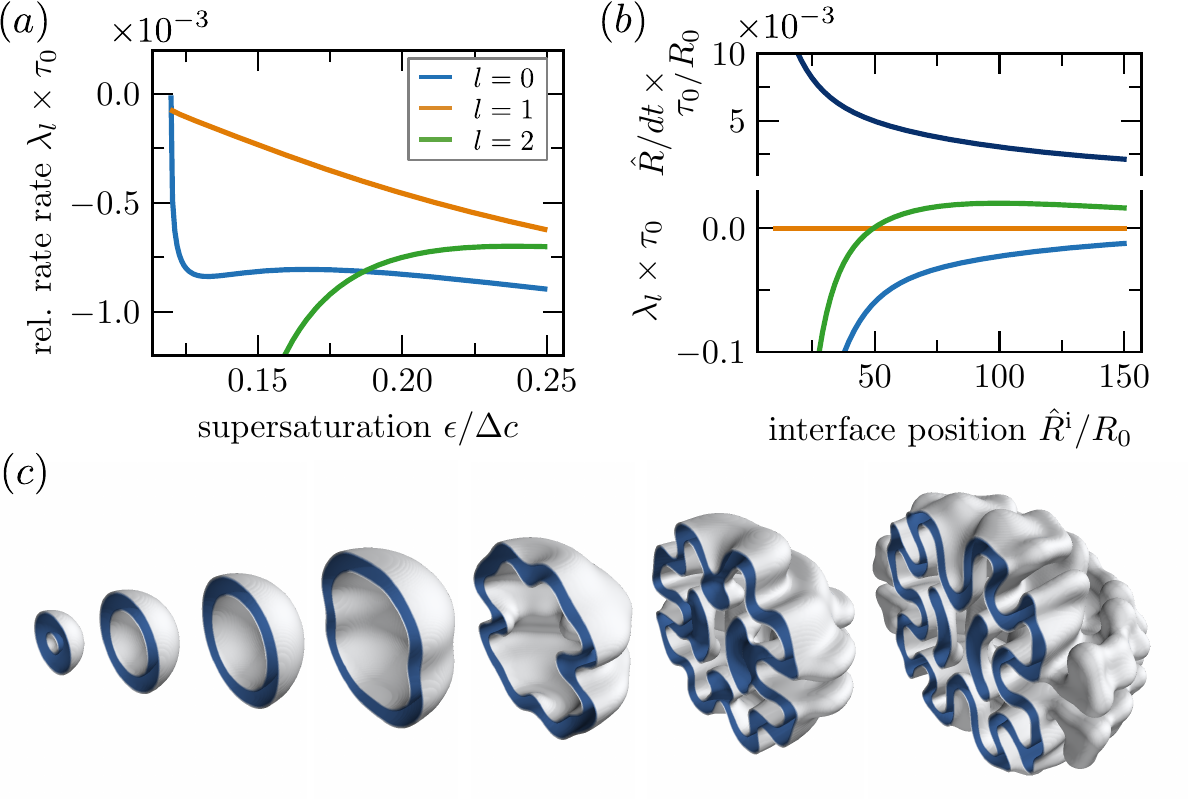}
\caption{(a-b) Relaxation time scales $\lambda_l$ of the first three perturbation modes ($l=0,1,2$) of a spherical symmetric stationary shells in a finite system with $R_s/R_0=60$ (a) and a constantly growing shell, together with its interface velocity $d\hat{R}/dt$, in an infinite system (b) with $\epsilon=0.12$.
If not mentioned otherwise, parameters and units are other parameters and units are the
same as in \fig{vacstate}(a).
(c) Cuts through temporal consecutive three-dimensional contour surfaces of a growing shell, same parameters as in \fig{model}(d).
\label{fig:shape_stab} }
\end{figure}

We now discuss the shape instability of droplets and shells by perturbing the interface radii as  $R = \hat{R} + \delta R(\theta,\varphi) $ and denoting the perturbed concentration profiles $c^i(r,\theta,\varphi) = \hat{c}^i(r) + \delta c^i(r,\theta,\varphi)$, in each domain $i$.
The dynamics of these perturbations can be decomposed to linear order in spherical harmonic Eigenmodes
\begin{gather}
    \delta R = \epsilon_l \text{Y}_{lm}(\theta,\varphi) \text{exp}(\lambda_l t)  \qc \\
    \delta c^i = \text{b}_l^i( r) \text{Y}_{lm}(\theta,\varphi) \text{exp}(\lambda_l t)  \qc
\end{gather}
where $\lambda_l$ is the relaxation rate of mode $l=0,1,2,\dots$, $\epsilon_l$ its amplitude, $\text{Y}_{lm}(\theta,\varphi)$ denote spherical harmonics, and $\text{b}_l^i( r)$ are combinations of modified spherical Bessel functions satisfying the appropriate boundary conditions of each domain, see App.~\ref{app:stationary_sol_pert}. 
The boundary condition~(\ref{eq:bc_inter}) becomes 
\begin{gather}
\delta c^i(\hat{R}) = \left(\beta \gamma \delta H(\hat{R})/\delta R -\partial_r \hat{c}^i(\hat{R})\right) \delta R \qc
\end{gather}
where  $\delta H /\delta R = [l(l+1)-2]/R^2 $ for a perturbed droplet. 
The linearized interface dynamics reads
\begin{gather}
\label{eq:pert_int_vel}
    \frac{d \delta R}{dt} = \frac{\mathbf{\delta j}^\ins(\hat{R}+\delta R) - \mathbf{\delta j}^\out(\hat{R}+\delta R)}{c^\ins(\hat{R}) - c^\out(\hat{R})} \mathbf{e}_r \qc 
\end{gather}
where $\delta \mathbf{j}^i(\hat{R}+\delta R)  \mathbf{e}_r= -D^i \left( \partial^2_r \hat{c}^i(\hat{R})\delta R +\partial_r \delta c^i(\hat{R}) \right)$.
For a stationary spherical droplet, the mode $l=2$ can become unstable with $\lambda_l>0$.
The stability diagram is shown in \fig{vacstate}(c) and (d) for two values of the ratio $D^\den/D^\dil$ of diffusion coefficients as a function of supersaturation $\epsilon$ and turnover $\eta$.
Stable stationary droplets exist in the blue region. As the supersaturation is increased, an $l=2$ instability occurs on the boundary to the green region. Inside the green region, spherical droplets are unstable, elongate their shape and typically divide in three dimensions~\cite{zwicker:2017a}.
Stable shells occur in regions shaded red. For parameter values used in \fig{vacstate}(c), spherical shells only occur when spherical droplets have already become unstable (region shaded both green and red). Therefore, in this region, vacuole solutions are only found if initial conditions are carefully chosen.
In contrast, for the parameters used in \fig{vacstate}(d), shell solutions are robustly 
found within a parameter region (red). In the region that is shaded in both red and blue or green, stable shells exist, but their formation depends on initial conditions.
The vertical red dashed lines indicate the spinodal concentration in the dilute phase. Supersaturation values beyond this line lead to a spinodal instability in the outside domain, and the single droplet or shell state does no longer exist.

\begin{figure}
\centering
\includegraphics[width=0.5\textwidth]{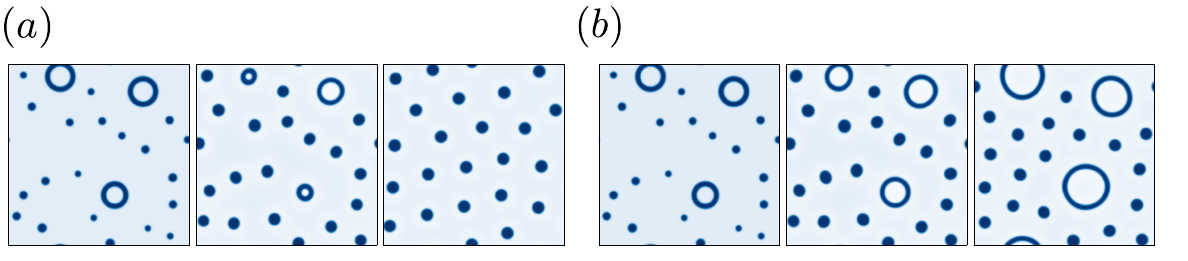}
\caption{
Time courses of emulsion after randomized initialization (see App.~\ref{app:numerics}) of droplets and spherical shells in a two-dimensional system for $\epsilon = 0.098$ (a) and $\epsilon = 0.104$ (b) with $\eta=0.001$, all other parameters identical as in \fig{model}(c).
\label{fig:emu}}
\end{figure}

In a finite system, stable spherical shells do not undergo a shape instability because the relaxation rates $\lambda_l$ are always negative. \fig{shape_stab}(a) shows the relaxation rates of modes $l=0,1,2$ for a
stationary shell.
In an infinite system, shells grow in size. Such growing shells can, beyond a critical size, undergo a shape instability with $\lambda_2>0$, see \fig{shape_stab}(b).
This is similar to the Mullins-Sekerka instability of constantly growing passive droplets in a supersaturated environment~\cite{mullins:1963b}.
The temporal evolution of a growing shell is shown in \fig{shape_stab}(c). After an initial growth phase, the spherical shell becomes unstable and folds into a shell with a complex morphology. This shell further grows and subsequently splits, generating elongated droplets inside the shell.

\emph{Emulsions of shells and droplets} - 
When we initialize a large box with a random set of droplets and shells, the system relaxes to a steady state. Two examples of time courses are shown in \fig{emu}(a,b).
For parameters where droplets but not shells exist as stable stationary solutions, shells shrink and turn into droplets, while droplets reach a stationary size, see \fig{emu}(a).
For parameters where droplets and shells can coexist, the system reaches a steady state of an emulsion of droplets of specific size and shells of variable size, see \fig{emu}(b). Note that in both \fig{emu}(a) and (b), Ostwald ripening is suppressed~\cite{zwicker:2015}.

\emph{Conclusion} - 
We have shown that spherical shells can generically exists in a minimal model of a chemically active  mixture. These shell states are stable in a broad range of values of supersaturation $\epsilon$ and turnover $\eta$.
While active droplets reach a characteristic size set by kinetic parameters, spherical shells grow until they fill the available space. 
Strikingly, both droplets and shells can undergo shape instabilities. Stationary droplets typically elongate and divide (in three dimensions), while shells only become unstable when they grow, giving rise to stable liquid shells.

The mechanism of shell formation relies on a spinodal instability inside the droplet resulting in a shell with  a dilute phase inside and outside.
In passive binary mixtures, a shell is not thermodynamically stable because the Gibbs-Thomson relation \eq{bc_inter} can not be satisfied at both interfaces at constant chemical potential. 
However, in multi-component systems, three-phase coexistence can occur, which allows for shells to coexist with two other phases~\cite{gibbs:1876,mao:2019a,mao:2020}.
The active shells described here are non-equilibrium structures that maintain chemical and diffusion fluxes at steady state.

Our theoretical work suggests that the formation of shells is facilitated if the dilute diffusion coefficient is bigger than the diffusion coefficient in the dense phase. Recently, liquid shells were observed in synthetic active emulsions, where $D^\dil \approx 50 \, D^\den$~\cite{bergmann:2023}, providing support for this result. Moreover, in such experiments, coexistence of active droplets and active shells was observed. 

Our work reveals that a minimal model of chemically active droplets has a rich phenomenology giving rise to emulsions that can suppress ripening and exhibit droplet division together with stable and unstable liquid shells.
Our work reveals the rich dynamical behavior of active emulsions
exhibiting the coexistence of active droplets and shells, and generating complex morphologies and topologies of liquid phases. 
Our work opens a new avenue to discover and classify the rich topological features in chemically-active emulsions.

\bibliography{vac}

\newpage
\onecolumngrid
\appendix
\section{Mobility and reaction rates in the continuous model}
\label{app:mobility}
We choose a concentration dependency of the mobility coefficient as 
\begin{equation}
    \Gamma(c) = \Gamma^\dil+ (\Gamma^\den-\Gamma^\dil) \left[1+ \tanh\left(\frac{c-(c^\dil+c^\den)/2}{\omega_c} \right)\right] \qc
\end{equation}
with $\omega_c$ describing the concentration range over which the mobility varies across the interface. If $\omega_c\ll \Delta c$, where, as in the main text  $\Delta c = c^\den- c^\dil$, the mobilities are almost constant within the two phases. In this limit, the diffusivities $D(c)=\Gamma(c) \partial f^2(c)/\partial c^2$ approaches $D^\dilden=b \Gamma^\dilden$ when evaluated at $c^\dilden$.

A third-order polynomial smoothly interpolates the reaction flux across the interface area. We define the interface region as concentrations within the bounds $c_B^\dilden = c^\dilden + \eqdelta /4$. For the choice of $c^\dil=0$ and $c^\dil=1$, the corresponding bounds, i.e. $c_B^\dil=0.25$ and $c_B^\den=0.75$, are within the spinodal regime, since the spinodal concentration in the dilute and dense phase read $c_\text{s}^\dil=0.211\dots$ and $c_\text{s}^\den=0.789 \dots$, respectively.

\section{Numerical methods}
\label{app:numerics}
We solve the continuous equations via spectral methods by using discrete Fourier transforms. When $\Gamma^\dil=\Gamma^\den$, the fourth-order term is linear and can easily be solved implicitly, while the remaining part of the Cahn-Hilliard equation \eq{dyn_eq} is non-linear.  We treat explicitly these terms by using an implicit-explicit Runge-Kutta scheme of second-third order \cite{ascher:1997}. For all other cases, i.e., $\Gamma^\dil \neq \Gamma^\den$, we use a standard explicit Runge-Kutta scheme of the third order.

All simulations were done in square boxes with $N=256$ grid points in each dimension for three-dimensional systems and $N=512$ grid points in each dimension for two-dimensional systems . We have chosen the system dimensions as $L=150 R_0$ for the three-dimensional systems and $L=300 R_0$ for the two-dimensional system. As in the main text, $R_0$ is a length scale associated to the interface width defined as $R_0 = 6 \beta \gamma /\Delta c$. Thus, there are always 3-4 grid points in the interface region.

For the random initialization of the emulsion of shells and droplets, we place twenty droplets with random radii sampled from a uniform distribution of  $R/R_0=6-9$ and three spherical shells with random radii sampled from $R^\Rio/R_0=7-14$ and random width $L/R_0=2-7$. We position these droplets and shells randomly in space. However, we ensure a minimal distance between the objects of 4 times their outer radius to other droplets or shells. We initialize the outside and core domain at the dilute concentration $c^\dil$, while we choose the inside domain of droplets and shells to be the dense concentration $c^\den$ initially.

\section{Stationary solutions}
\label{app:stationary_sol}
The stationary solutions of \eq{reac-dif} with the boundary conditions in the corresponding domains for the droplet state read
\begin{align}
    \phi^\ins(r) =& \left(\frac{\eta}{k^\den} + \frac{2 \beta \gamma}{R}\right)\frac{\text{i}_0(\Lambda^\den r)}{\text{i}_0(\Lambda^\den R)}+\frac{s^\den}{k^\den} \qc \\
    \phi^\out(r) =& \left(\frac{2 \beta \gamma}{R} -\epsilon \right) \frac{\text{k}_0(\Lambda^\dil r)+\frac{\text{k}_1(\Lambda^\dil R_s)}{\text{i}_1(\Lambda^\dil R_s)}\text{i}_0(\Lambda^\dil r)}{\text{k}_0(\Lambda^\dil R)+\frac{\text{k}_1(\Lambda^\dil R_s)}{\text{i}_1(\Lambda^\dil R_s)}\text{i}_0(\Lambda^\dil R)}+\epsilon + \phi_0^\dil \qc 
    \label{eq:statsol_out}
\end{align}
where $\Lambda^\dilden=\sqrt{k^\dilden/D^\dilden}$, $\text{i}_0(x) = \sinh(x)/x$ and $\text{k}_0(x) = \exp(-x)/x$ are the Modified Spherical Bessel Functions of the first and second kind of zeroth order, while $\text{i}_1(x) = (x \cosh(x) - \sinh(x))/x^2$ and $\text{k}_1(x) = \exp(-x)(x+1)/x^2$ are the Modified Spherical Bessel Functions of the first and second kind of first order. Here, we have considered a finite system with radius $R_s$, and assumed no-flux boundary conditions at the boundary of the the outside domain, i.e $r=R_s$. In the limit of infinite systems $R_s \rightarrow \infty$, the fraction $\text{k}_1(\Lambda^\dil R_s) / \text{i}_1(\Lambda^\dil R_s) \rightarrow 0$. Thus, only the term with the Modified Spherical Bessel Functions of the second kind $k_0(x)$ in \eq{statsol_out} remains relevant in this limit.
For the state of the spherical shell, we find
\begin{align}
    \phi^\core(r) =& -\left(\epsilon + \frac{2 \beta \gamma}{R^\Ri} \right)\frac{\text{i}_0(\Lambda^\dil r)}{\text{i}_0(\Lambda^\dil R^\Ri)}+\epsilon +\phi_0^\dil \qc \\
    \phi^\ins(r) =  &\frac{s^\den}{k^\den}
    +\Bigg( \frac{\text{i}_0( \Lambda^\den R^\Ri)}{\text{k}_0(\Lambda^\den R^\Ri)} - 
  \frac{\text{i}_0(\Lambda^\den R^\Ro)}{\text{k}_0(\Lambda^\den R^\Ro)} \Bigg)^{-1} \Bigg(\frac{\frac{\eta}{k^\den} - \frac{2 \beta \gamma}{R^\Ri}}{
 \text{k}_0(\Lambda^\den R^\Ri)} - \frac{\frac{\eta}{k^\den}  + \frac{2 \beta \gamma}{R^\Ro}}{ \text{k}_0(\Lambda^\den R^\Ro)} \Bigg) \text{i}_0(\Lambda^\den r) +
    \nn\\    
     +&\Bigg( \frac{\text{k}_0( \Lambda^\den R^\Ri)}{\text{i}_0(\Lambda^\den R^\Ri)} - 
  \frac{\text{k}_0(\Lambda^\den R^\Ro)}{\text{i}_0(\Lambda^\den R^\Ro)} \Bigg)^{-1} \Bigg(\frac{\frac{\eta}{k^\den} - \frac{2 \beta \gamma}{R^\Ri}}{
 \text{i}_0(\Lambda^\den R^\Ri)} - \frac{\frac{\eta}{k^\den}  + \frac{2 \beta \gamma}{R^\Ro}}{ \text{i}_0(\Lambda^\den R^\Ro)} \Bigg) \text{k}_0(\Lambda^\den r) \,.
\end{align}
The solution of the outside domain, $\phi^\out(r)$,  is identical to \eq{statsol_out} when $R$ is replaced with $R^\Ro$.

\section{Stationary solutions of the perturbations}
The corresponding boundary conditions lead to the solutions
\label{app:stationary_sol_pert}
\begin{align}
    b_l^\ins(r) =  \epsilon & \left(\frac{\beta \gamma [l (l + 1) - 2]}{\hat{R}^2} -\partial_r \hat{\phi}^\ins(\hat{R}) \right) \frac{\text{i}_l(\Lambda_l^\ins r)}{\text{i}_l(\Lambda_l^\ins \hat{R})}\qc \\
    b_l^\out(r) =\epsilon & \left(\frac{\beta \gamma [l (l + 1) - 2]}{\hat{R}^2} -\partial_r \hat{\phi}^\out(\hat{R}) \right) \frac{\text{k}_l(\Lambda_l^\out r)-\frac{\text{k}_l'(\Lambda_l^\out R_s)}{\text{i}_l'(\Lambda_l^\out R_s)}\text{i}_l(\Lambda_l^\out r)}{\text{k}_l(\Lambda_l^\out \hat{R})-\frac{\text{k}_l'(\Lambda_l^\out R_s)}{\text{k}_l'(\Lambda_l^\out R_s)}\text{k}_l(\Lambda_l^\out \hat{R})} \qc \label{eq:statpertoutbessel}
\end{align}
where $\text{i}_l(x)$ and $\text{k}_l(x)$ are the Modified Spherical Bessel Functions of the first and second kind of $l$'th order. Furthermore, we have used $\text{i}'_l(x) = \partial_x \text{i}_l(x)$ and $\text{k}'_l(x) = \partial_x \text{k}_l(x)$, and we defined the inverse length-scales $\Lambda^i_l = \sqrt{(k^i-\Lambda_l)/D^i}$. 
Again, we have assumed no-flux boundary conditions at $R_s$ and in the outside domain. Similarly, to \eq{statsol_out}, the Modified Spherical Bessel Functions vanish in \eq{statpertoutbessel} for $R_s \rightarrow \infty$.
For the shell state, we find
\begin{align}
    b_l^\core(r) =& \epsilon_l^\Ri \left(-\frac{\beta \gamma [l (l + 1) - 2]}{(\hat{R}^\Ri)^2} -\partial_r \hat{\phi}^\text{cor}(\hat{R}^\Ri) \right) 
    \frac{\text{i}_l(\Lambda_l^\text{cor} r)}{\text{i}_l(\Lambda_l^\text{cor} \hat{R}^\Ri)}\qc \\
    b_l^\ins(r) =& \frac{ \Big[\text{k}_l(\Lambda_l^\ins R^\Ri) C^\Ro - \text{k}_l(\Lambda_l^\ins R^\Ro) C^\Ri \Big] \text{i}_l(\Lambda_l^ r) }{\text{i}_l( \Lambda_l^\ins  R^\Ro) \text{k}_l (\Lambda_l^\ins  R^\Ri) -  \text{i}_l( \Lambda_l^\ins  R^\Ri) \text{k}_l(\Lambda_l^\ins  R^\Ro)} +
    \frac{\Big[ \text{i}_l( \Lambda_l^\ins  R^\Ro) C^\Ri - \text{i}_l(\Lambda_l^\ins  R^\Ri) C^\Ro\Big] \text{k}_l(\Lambda_l^\ins r)}{\text{i}_l( \Lambda_l^\ins  R^\Ro) \text{k}_l (\Lambda_l^\ins  R^\Ri) -  \text{i}_l( \Lambda_l^\ins  R^\Ri) \text{k}_l(\Lambda_l^\ins  R^\Ro)} \qc
     \\
    C^\Ri =& \epsilon_l^\Ri\left(\frac{-\beta \gamma [l (l + 1) - 2]}{(\hat{R}^\Ri)^2} -  \partial_r \phi^\ins(\hat{R}^\Ri) \right) \qc \\
    C^\Ro =& \epsilon_l^\Ro\left(\frac{\beta \gamma [l (l + 1) - 2]}{(\hat{R}^\Ro)^2}  -  \partial_r \phi^\ins(\hat{R}^\Ro) \right) \qd
\end{align}
The outside domain for the shell state is identical to \eq{statpertoutbessel} when $\hat{R}$ is replaced by $\hat{R}^\Ro$ and $\epsilon_l$ with $\epsilon_l^\Ro$.

For the perturbation of a droplet, the strength of the perturbation $\epsilon_l$ appears in every term in \eq{pert_int_vel}. Thus, this equation directly determines the relaxation rate $\tau_l$. However, for the perturbation of the spherical shell,
$\epsilon_l^\Ri$ and $\epsilon_l^\Ro$ can not be chosen independently, because \eq{pert_int_vel} has to be fulfilled on both interfaces $R^\Rio$. These two constraints determine the ratio $\epsilon_l^\Ri/\epsilon_l^\Ro$ and $\tau_l$. 

\section{Shell width in the large system limit}
\label{app:Linfty}
In the infinite system, both interfaces move toward infinity. Thus, the curvature of both interfaces vanishes, and the system becomes effectively one-dimensional. Therefore, we can estimate the shell width $L^\infty$ by the size of a one-dimensional droplet in an infinite system. Assuming the droplet center at $x=0$, the stationary solutions of \eq{reac-dif} read
\begin{gather}
    \phi^\ins(x) = \frac{\eta}{k^\den}\frac{\cosh(\Lambda^\den x)}{\cosh(\Lambda^\den R)} +\frac{s^\den}{k^\den}, \;\text{for } |x|<R, \\
 \phi^\out(x) = -\epsilon \frac{\exp(\pm \Lambda^\dil x)}{\exp(\pm\Lambda^\den R)}+\epsilon +\phi_0^\dil, \; \text{for } |x| > R.
\end{gather}
The two interfaces at $\pm R$ are stationary if $D^\den \partial_x \phi^\ins(R) = D^\dil \partial_x \phi^\out(R)$, thus the interface velocity \eq{drop_vel} vanishes. Due to the absence of any Laplace pressure in 1D, there is no nucleation barrier, and the only solution of $R$ fulfilling the last condition reads
\begin{equation}
    R = \sqrt{\frac{D^\den}{k^\den}}  \text{arctanh}
  \left(\sqrt{\frac{D^\dil k^\dil}{D^\den k^\den}} \frac{k^\den \epsilon} {\eta} \right) \qc
\end{equation}
leading to the expression stated in \eq{Linfty}.
\end{document}